# Features and Operation of an Autonomous Agent for Cyber Defense

Michael J. De Lucia[1], Allison Newcomb[2], Alexander Kott[3]
U.S. Army Research Laboratory
2800 Powder Mill Road, Adelphi, Maryland, USA
301-394-0798

{michael.j.delucia2.civ@mail.mil , elizabeth.a.newcomb9.civ@mail.mil , alexander.kott1.civ@mail.mil}

## Abstract

An ever increasing number of battlefield devices that are capable of collecting, processing, storing, and communicating information are rapidly becoming interconnected. The staggering number of connected devices on the battlefield greatly increases the possibility that an adversary could find ways to exploit hardware or software vulnerabilities, degrading or denying Warfighters the assured and secure use of those devices. Autonomous software agents will become necessities to manage, defend, and react to cyber threats in the future battlespace.

The number of connected devices increases disproportionately to the number of cyber experts that could be available within an operational environment. In this paper, an autonomous agent capability and a scenario of how it could operate are proposed.  The goal of developing such capability is to increase the security posture of the Internet of Battlefield Things and meet the challenges of an increasingly complex battlefield. This paper describes an illustrative scenario in a notional use case and discusses the challenges associated with such autonomous agents.  We conclude by offering ideas for potential research into developing autonomous agents suitable for cyber defense in a battlefield environment.

## Motivation and Context

The Internet of Battlefield Things (IoBT) is "a set of interdependent and interconnected entities (e.g., sensors, small actuators, control components, networks, information sources)" that are composed and connected dynamically to support the goals of a military mission (US Army Research Laboratory, 2017; Kott et al., 2016).  These "things" will function with varying levels of autonomy in order to adapt to a broad range of mission goals and environments.  The sheer number of these things will far exceed the number of humans available to oversee their operation.  As a result, the things within the IoBT will require the support of autonomous agents, particularly for the purposes of their cyber defense.

The battlefield is a highly dynamic and uncertain environment, often dominated by adverse conditions degrading the effectiveness of communications and information networks that are the Warfighters' critical tools.  Battlefield networks are necessarily mobile and are composed of many heterogeneous devices.  Mobility and adversary actions result in topologies that change quickly and frequently.  This requires reestablishing connections as configurations change.  The lack of infrastructure in battlefield environments severely constrains the amount of bandwidth and computational capabilities available to Warfighters. Reliable, timely communication of accurate information is critical to the successful execution of every mission, but the resource constraints inherent to tactical networks threaten the delivery and assurance of vital information.

As an example, consider a robotic vehicle, an entity within the IoBT that gathers information for dismounted Warfighters. This unmanned vehicle collects and transmits images of buildings and roads as well as meteorological and geographic data. If a specific threat is detected by the robotic vehicle, the transmission of such information to the Warfighters takes priority over all other communication. The adversary may attempt to deploy a malware on the robotic vehicle in order to deny or degrade the vehicle's high-priority communications. An autonomous cyber agent—a software agent—resides on the robotic vehicle, senses the status of its environment, and chooses the best course of action to protect it. In this hypothetical example, the agent recognizes a software module's attempt to connect to a suspicious web site, one that is associated with downloaded malware. Blocking traffic between the robotic vehicle and that site is a course of action the agent chooses as a means to defend the vehicle.

Here, a few words about the terminology are in order. Autonomous agents can be a physical agent or a software agent. Autonomous agents can be physical entities, such as a robot or drone functioning independently and without direct human guidance; and they can be software entities, such as a cyber-defense agent. The focus of this article is software agents, specifically cyber defense agents. In the preceding example, the robotic vehicle is a physical autonomous agent. In addition, one or multiple software agents, such as a cyber defense agent, reside on a physical agent. An autonomous software agent can be defined as software which acts on its own without human intervention (Shiffman, 2012) or as a "self-activating, self-sufficient, and persistent computation" (Shattuck, 2015). For the purposes of this paper, the term "agent" is used to refer to software agents, not physical agents. Furthermore, we use the term "agent" to refer to a specific type of a software agent—an agent that specializes in cyber defense of things within the IoBT.

The remainder of the paper is organized as follows. In the next section, we elaborate on the argument that the enormous size of the future IoBT, along with the sophisticated cyber threats of the future battlefield, will necessitate the wide use of autonomous intelligent cyber defense agents. We then present a notional, illustrative scenario—a few moments in the life of a cyber defense agent. This is followed by a discussion of challenges and research topics that arise from considering the scenario.

## Autonomous Cyber Defense Agents Are a Necessity

Technologies such as "machine intelligence and networked communications" have spurred the growth and acceptance of connecting cell phones and other personal devices with everything from household appliances to automobiles. Similarly, the military is applying computational intelligence to its interconnected battlefield devices to make them smarter and thereby, more useful to Soldiers. The IoBT encompasses various sensors, vehicles, communication devices, computers, and information sources. It is certain that the future battlefield will be densely populated with a variety of interconnected devices (Kott et al., 2016).

To gain an appreciation for the size and complexity of the IoBT, consider the following:
It has been estimated (Anonymous, 2017) that the commercial or consumer Internet of Things (IoT) grew from 2 billion devices in 2006 to 15 billion devices in 2015. The same source estimates that by the year 2020, 200 billion devices will populate the IoT. If the proliferation of IoBT (military devices) follows the growth of IoT (consumer devices), it is clear that human Warfighters, will require an augmentation by autonomous cyber defense agents to monitor and defend battlefield devices.

Additional complexity becomes apparent when one considers that the IoBT will have to operate effectively within environments that it neither owns nor controls. For example, a military force may be operating within a city where the majority of computing devices—the consumer devices—belong to the neutral civilians but are also potentially controlled by the adversary. Furthermore, in the case of IoBT the adversary will actively pursue compromise, capture, or destruction of the battlefield devices.

Given that cyber-attacks will occur frequently and at a high pace that will surpass human ability to respond in a timely fashion, decisions on the most appropriate course of cyber defense actions will have to occur in near real-time.

## An Illustrative Operating Scenario

In order to illustrate how an autonomous cyber defense agent might operate, we offer a notional operating scenario. In this scenario, Blue refers to friendly forces and Red refers to the adversary. Blue-17, Blue-19, and Blue-23 are peer cyber defense agents. Each agent is installed by a human operator on its respective device within the Blue IoBT (e.g., an Android phone) and is tasked with cyber defense of that device. Blue-C2 is the Command and Control (C2) node that commands, coordinates, and supports all other Blue agents, at least when communications between an agent and the Blue-C2 node are available. There is only one Red agent—Red-35—in our simple scenario.

The protagonist of our scenario is Blue-17, a cyber defense agent that has been installed on a friendly device; it continuously monitors Blue space network and scans event logs looking for suspicious activity. The antagonist is Red-35, a malware agent successfully deployed by the Red forces on the device defended by Blue-17. The events unfold, briefly, as follows.

Blue-17 detects a hostile activity associated with Red-35 and attempts to contact the Blue-C2 for additional remediation instructions. Unfortunately, the communications are heavily contested by the adversary, and response from Blue-C2 is not coming. Therefore, Blue-17 decides to contact peer agents (Blue-19 and Blue-23) in search for relevant information. Although Blue-19 and Blue-23 receive this message from Blue-17, their responses are not arriving to Blue-17. Having heard nothing within a reasonable waiting time, Blue-17 independently formulates and executes a set of actions to defeat Red-35. However, having completed these actions, Blue-17 receives a belated reply from Blue-23. Blue-17 determines that Blue-23 is compromised because the response is suspicious. Given the extreme seriousness of this situation, Blue-17 neutralizes Blue-23 and places a copy of itself on the device that was being protected by Blue-23.

Table 1 provides a hypothetical timeline of these events and the agents' actions. Durations are intended to merely illustrate the flow of time in the scenario and are in no way representative of execution speeds of any actual hardware or software. Following the table, we discuss each step of the scenario in more detail.

| Step | Elapsed Time | Condition/Event | Active Software Agent | Action |
|---|---|---|---|---|
| 1 | H = 0 sec | Start up | Blue-17 | Monitor network traffic and scan logs |
| 2 | H = H + 0.100 sec | Hostile software agent compromises device and network | Red-35 | Red-35 infiltrates Blue device and network. Blue-17 does not notice the infiltration. |
| 3 | H = H + 0.200 sec | Red-35 begins operations. Suspicious activity detected | Red-35 and Blue-17 | Red-35 conducts malicious activities. Blue-17 detects an activity and predicts probable compromise. |
| 4 | H = H + 0.22 sec | Compromise suspected | Blue-17 | Contacts C2 node |
| 5 | H = H + 3.00 sec | No response from C2 node | Blue-17 | Contact Blue-19 and Blue-23 agents |
| 6 | H = H + 5.00 sec | Message among Blue peer agents | Blue-19 and Blue-23 | Receive message from Blue-17 |
| 7 | H = H + 10.00 sec | Message acknowledgement time out | Blue-17 | Choose alternate course of action |
| 8 | H = H + 12.00 sec | No communication with peer defensive agents | Blue-17 and Red-35 | Block or redirect Red-35 communication. Red-35 is unable to defend itself. |
| 9 | H = H + 23.00 sec | Response received from Blue-23 | Blue-17 | Blue-17 determines that the response is invalid |
| 10 | H = H + 28.00 sec | Neutralize compromised Blue agent | Blue-17 | Quarantine or destroy Blue-23 software code |
| 11 | H = H + 28.3 sec | Replicate and overwrite | Blue-17 | Copy to device |

Table 1. Hypothetical Timeline of Agents' Actions.

**Scenario Steps 1–2**
In the scenario, Blue-17 passively monitors the inbound and outbound network communications using a lightweight intrusion detection system (IDS) such as FAST-D (Yu and Leslie, 2018). FAST-D is a software that performs intrusion detection using far less computational resources than alternative solutions. Its algorithm uses hash kernels and byte patterns as signatures to examine the packet payload content of all network communications. Additionally, Blue-17 scans the device logs looking for indicators of compromise (privilege escalation, abnormal crashes, failed logins, etc.).

**Scenario Steps 3–4**
Blue-17 sends a message to its C2 node for further remediation instructions and verifications. A C2 node is one that is central (root) and is responsible for the management and tracking of all Blue agents. A C2 node resides in a central location that may be the tactical operations center. The message sent to the C2 node is encrypted to protect the confidentiality and integrity and is in a predefined format for agent messages. This message is split up into many small segments, is blended into normal traffic to masquerade as other legitimate traffic, and sent through different routes within the network in order to avoid an attacker from intercepting or detecting the agent message sent to the C2 node. Lastly, the address of the C2 node changes over time based on a deterministic algorithm, known to all agents to make it more difficult for Red-35 to discover its location.

**Scenario Step 5**
After some reasonable waiting time passes, and Blue-17 does not receive a reply back from the C2 node, it decides as an alternative action to send out a request to its peer agents (Blue-19 and Blue-23) for their remediation recommendations. Again this message is sent out using an encrypted predefined format for agent messages as previously described in sending a message to the C2 node. The message is sent directly to the peers and is blended into other network traffic. The peer agents are neighbors to Blue-17 and are also be under the management of the C2 node.

**Scenario Step 6**
Both Blue-19 and Blue-23 have received the message from Blue-17. After some delay, Blue-23 sends a response and recommendation back to Blue-17 using the same method for sending a message to a peer agent.

**Scenario Steps 7–8**
Within a specified time interval, Blue-17 has not received a response from either its C2 node or its peers (Blue-19 and Blue-23). Blue-17 requested further verification of the threat before taking a destructive action against Red-35. However, since a response was not received, Blue-17 decides to take action on the perceived Red-35 malware agent threat. The Blue-17 agent first isolates the Red-35 malware agent and its communication in a honeypot to observe the actions taken by the attacker. Blue-17 has taken this action since it is not confident in its assessment of the detection of the perceived Red-35 agent.

**Scenario Step 9**
After some time has passed, and Blue-17 has already taken action, a response from Blue-23 is received. Blue-23's response contains a signature and timestamp that allows Blue-17 to determine the authenticity of the message received. However, as Blue-17 verifies the response message from Blue-23, it determines that the message signature is not valid and rejects the message. Blue-17 concludes that Blue-23 may be compromised.

**Scenario Steps 10–11**
Blue-17 has discovered that Blue-23 has been compromised. Blue-17 takes action to quarantine Blue-23. Blue-17 clones itself to create a pristine copy of the defensive agent. Blue-17 initiates the overwriting of the Blue-23 agent image with a fresh copy of a defensive agent with the initial state of Blue-17. The agent package is sent via an encrypted message from Blue-17 to the container management of Blue-23. The container management package of the agent uses cryptographic authentication, allowing the overwriting to occur. Blue-23 is restored back to a fresh agent image and is no longer infected.

## Discussion of Challenges and Requirements

Having offered a scenario—simple yet sufficiently illustrative of potential difficulties—we now have a basis for discussing the technical challenges and requirements. One of the requirements illustrated in part by the scenario is that a defensive agent must reside outside of the operating system of the device it is protecting. This arrangement avoids the possibility of the malware providing false information or changing the view of the defensive agent (i.e., Blue-17). Malware can disable processes or deceive (e.g., by providing false information) software such as the anti-virus (AV) software or firewall on a device (Baliga et al., 2007). A logical separation at the hardware level between the operating system being protected and the defensive agent will protect the Blue-17 agent from being compromised by malware infection. The defensive agent will require access in a secure manner to all of the files and state from its outside view, while being protected from any threats affecting the Blue-17 operation or integrity.

Additionally, because the Blue-17 agent resides outside of the protected operating system, Red-35 will not be able to detect Blue-17's presence or any of its actions. A traditional placement alternative for an agent that resides outside of the protected operating system, would be a distributed or network-based sensor. That configuration comes with a tradeoff: an agent (Blue-17) at the network level would not be able to monitor the file system of the protected operating system. Therefore, the Blue-17 agent must reside on the same physical device outside of the operating system being protected.

Also, in order for the agent to move around freely among the devices within the protected network, the agent must be unconstrained by any particular operating system. It is also presumed that the container in which the agent runs has been pre-installed on the device to which agents can migrate freely to, such as in the case with Blue-17 overwriting Blue-23.

Clearly required, as illustrated in our scenario, is a fast, highly reliable and low-resource means of detecting potentially malicious activity. For example, using a low-resource intrusion detection software, Blue-17 was able to detect rapidly and with a significant degree of assurance a suspicious activity performed by a sophisticated agent Red-35. Additional solutions could be employed that use supervised machine-learning approaches, coupled with features such as network packet inter-arrival times, packet sizes, Transport Control Protocol (TCP) flags, and such, to perform detection of malware infiltration. However, in either case it is important to understand the limitations (i.e., inability to detect malware within encrypted communications) of the intrusion detection algorithm chosen to perform detection of malicious communications. It is also important to know the possible ways an attacker could evade (fragmentation attack, encrypted attack, etc.) the IDS. Successful evasion by an attacker will result in a missed attack, also called a false negative. It is also critical for an autonomous agent employing an IDS algorithm to have a low false-positive rate (misclassified legitimate traffic as an attack) and low false-negative rate (missed attack). In a military context a false-positive in an autonomous cyber defense agent will impact the mission by denying legitimate and essential communication.

Another challenging requirement is the need to manage the degree of the agent's autonomy. Blue-17 could be fully autonomous or semiautonomous. In our scenario, Blue-17 is fully autonomous, as defined by the lack of human intervention at any point. Consequently, Blue-17 must be highly confident in the detection event and its resultant course of action. The agent's actions must avoid any adverse reaction, such as degrading network performance or dropping nodes on the network as a mitigation, resulting in access denials. Alternatively, Blue-17 could act as a semiautonomous agent, with varying levels of interaction between the agent and human controllers, which present many challenges of their own (Kott and Alberts, 2017). For example, Blue-17 could detect a potential compromise and then defer to a human analyst (e.g., by contacting the C2 node and waiting for instruction) in a case where there is low to moderate confidence in the detection event.

The agent will require the ability to share threat data directly with its peers (e.g., Blue-17 had to share data with Blue-23 and Blue-19) and orchestrate coordinated defensive actions when necessary. Additionally, the agent must also be able to work in an isolated environment and make appropriate decisions independently, as Blue-17 had to do when it failed to receive response from either Blue-C2 or Peer's agents. These agents will need to store pertinent information on detected attacks and outcomes (successful vs. unsuccessful) of the selected mitigation strategies. This information will need to be stored in a compressed format due to the limited resources characteristic of the various devices of IoBT. On the other hand, when the agents return to a less-contested environment where power and bandwidth are less constrained and more reliable, the data would be uploaded to a central repository. Lessons learned (quantitative measures of outcomes) and specifics on detected attacks would be compiled to improve the process of informing other autonomous agents. This arrangement would expand and enrich the agents' knowledge and ability to learn from historical decision-making strategies.

The agent (i.e., Blue-17, Blue-19, or Blue-23) hosted within the IoBT environment must process and synthesize the information it produces or receives from other agents to a subset relevant to the Warfighters' cognitive needs (Kott et al., 2016). For example, of all the alerts produced by the agents, the Warfighter will only need to be aware of a small subset to form a situational awareness of on-going cyber-attacks. This filtered information must be relevant and trustworthy to both the IoBT device and the Warfighter's cognitive needs. Providing incorrect or irrelevant information could cause significant and negative impact to the mission (Kott et al., 2016). Further, information stored by agents on IoBT devices must be distributed and obscured from the adversary. An approach to secure the distributed agent information within an IoBT environment is to split the data into fragments and disperse them among the many devices in a way to thwart the adversary's ability to reconstruct the information based on a number of captured segments (Kott et al., 2016a).

Ideally, the agents' performance would be evaluated in order to refine and share successful strategies with other agents. Performance in this context includes the agents' decision-making value, timing, and the resulting impacts of the courses of action executed (e.g., Blue-17 was successful—what factors contributed to these successes?). This supports the need for agents to be able to learn from their actions as well as the actions of other agents via machine-learning techniques.

The agent could employ a combination of supervised and unsupervised machine learning. The lessons learned and outcomes of the course of action taken by an agent could be used with a reinforcement-based machine-learning algorithm. For example, the successful course of action executed by Blue-17 with respect to defeating Red-35 would receive a positive reward. This approach could be used to expand the knowledge of the autonomous agents, thereby improving the agents' performance and effectiveness.

Another requirement of these agents will be trust management between devices. Each device on the network will require software-based logic to participate in the network with a full degree of trust and access. This logic can be preinstalled or can be acquired from a peer node by a device that seeks to join the network in a comply-to-connect mode of operation. Once compliance conditions are met, the agent can be transferred to other network member nodes. For example, in our scenario, Blue-17 needed a way to determine that Blue-23 is no longer trustworthy. At the same time, Blue-17 had to elicit a sufficient degree of trust from the node where Blue-23 resided in order to overwrite the Blue-23 image.

Device-to-device transfer of the agents—such as the move of a copy of Blue-17 to the node originally defended by Blue-23—necessarily raises concern for unintended propagation and behaviors beyond the intended network, as witnessed with the Morris worm (Qing and Wen 2005; Spafford 1989) and the more recent Stuxnet attack (Farwell and Rohozinski, 2011). Findings from studies on limiting the spread of malware in mobile networks (Zyba et al., 2009; Li et al., 2014) could be adapted to manage the propagation of defensive agents. Another potential solution to controlling propagation is to require consensus approval of a certain number of nodes prior to enabling transfer of the agent to a new device. A suggested approach is to define boundary rules to determine whether the agent has been transferred outside its intended network. When the boundary rules evaluate to a true condition (out of bounds), mandatory removal of the agent or a self-destruct sequence would be triggered. The effects of these combined approaches to controlling propagation require additional research.

While autonomous agents should be free to learn, act, and propagate, careful thought should be given to methods that would constrain behaviors within the bounds of legal and ethical policies, as well as the chain of command. For example, it would be undesirable if Blue-17 were to learn that requests to Blue-C2 are generally fruitless and should not be attempted. An agent that is fully autonomous must be able to operate within an appropriate military C2 construct (Kott and Alberts, 2017). It is imperative that a software agent be bounded in its propagation, yet capable to move around freely between authorized devices.

# Conclusions

With a large number of devices within the future IoBT it will be imperative for these devices to be able to defend themselves. Further, personnel who interact with the IoBT devices will not be cyber security experts and will be focused on the execution of the mission without the ability to continuously monitor the health of their devices. Autonomous cyber defense agents will be required to augment and multiply military forces. Such agents will need to possess awareness of, and ability to learn about, threats in near real-time. The agents will need to have the capability to reliably and predictably self-propagate, sense malicious activity, and disseminate information among trusted network members.